\documentclass[11pt]{article}
\usepackage{epsfig}
\usepackage{natbib}
\usepackage{amssymb,amsmath}

\newcommand{\ups}{\rule{0pt}{15pt}}

\title{\bf \Large The statistics of the photometric accuracy based on MASS data and the evaluation of high-altitude wind}

\author{ \bf V. Kornilov\footnote{E-mail: victor@sai.msu.ru}}

\date{\it Sternberg astronomical institute, Universitetsky pr-t, 13, Moscow, Russia}
\begin{document}
\maketitle
\begin{center}
\small Accepted for publication in Astronomy Letters
\end{center}
\bigskip

\begin{abstract}
The effect of stellar scintillation on the accuracy of photometric measurements is analyzed. We obtain a convenient form of estimaton of this effect in the long exposure regime, when the turbulence shift produced by the wind is much larger than the aperture of the telescope. A simple method is proposed to determine index $S_3$ introduced by \citet{ken2006}, directly from the measurements with the Multi Aperture Scintillation Sensor (MASS) without information on vertical profile of the wind. The statistics $S_3$ resulting from our campaign of 2005 -- 2007 at Maidanak observatory is presented. It is shown that these data can be used to estimate high-altitude winds at pressure level 70 -- 100 mbar. Comparison with the wind speed retrieved from the NCEP/NCAR global models shows a good agreement. Some prospects for retrieval of the wind speed profile from the MASS measurements are outlined.
\end{abstract}



\section{Introduction}

The accuracy of measurements of brightness of astronomical objects is affected by many factors. It is clear that they are influenced in different ways depending on the applied photometric techniques. One of these effects is the stellar scintillation caused by {\it optical turbulence} (OT) in the Earth's atmosphere. Unlike other variations of the atmospheric transparency, the light fluctuations due to the scintillation do not correlate for the images separated by more than the isoplanatic angle ($2''-5''$ in the optical range). Therefore, errors introduced by this effect appear always and increase when the brightness of two stars is compared.

This problem was noted and studied a long time ago \citep{Young69, DravI, DravII, DravIII}. Recently, interest in this problem manifested itself again, but in the overall context of a study and characterization of the OT in various astronomical observatories and prospective sites. In the paper \citet{ken2006} a special index $S_3$ have been introduced to relate the vertical distribution of the turbulence and wind speed to the variations of the light flux as measured in a large telescope.

Monitoring of this parameter and its analysis is very useful for comparing different sites and for predicting or planning of photometric observations. However, the approach used in cited work implies prior restoration of the OT profiles from the MASS -- Multi Aperture Scintillation Sensor \citep{MASS, mnras2003} data and the use of additional information on wind speed. This is a very intensive procedure, besides burdened by additional errors

In this paper we propose a method to determine the index $ S_3 $ directly from the MASS measurements without using additional information. The presented method is preceded by a theoretical consideration of the scintillation in the temporal domain.

\section{Temporal averaging of the scintillation}

The approximation of weak perturbations implies the independence of the wavefront distortions produced in a turbulent layer from the turbulence located above it. This leads to the fact that the scintillation index $s^2$ --- the variance of relative fluctuations of the light intensity, is described by the sum of the scintillation indices produced by individual layers:
\begin{equation}
s^2 = \int_0^\infty C_n^2(h)\, W(h)\, {\rm d}h \quad,
\label{eq:is}
\end{equation}
where $W(h)$ is the {\it weighting function} (WF) which depends on the size and shape of the receiving aperture and does not depend on the altitude distribution of the structural coefficient of the refractive index $C_n^2(h)$. WF represents a power of the scintillation generated by a layer of unit intensity located at a height $h$.

The MASS method involves simultaneous measurement of the scintillation indices in 4 concentric apertures of different diameters, leading to 4 normal and 6 diffential scintillation indices. The vertical OT profile is restored from the measured indices and the theoretically calculated WFs \citep{mnras2003, MD2007}. The calculation of the set of functions $W(h)$ assumes that the light intensity measurement has a ``zero exposure'', i.e. the averaging factor is related only to the receiving aperture. This is not quite true in real measurements, so the problem of finite time of the measurement was considered separately by \citet{FexpT2002}.

The question of the quantitative effect of averaging on the scintillation was considered in many papers, see e.g. \citet{DravI}. We consider below this problem using the terminology of the MASS theory \citep{mnras2003, MD2007} and the results given by \citet{Martin1987} and \citet{FexpT2002} in order to obtain formulas more convenient for practical application.

It is obvious that the variance $\sigma^2$ of the relative fluctuations of the light intensity measured with finite exposure time is determined by the integrated effect of all turbulent layers on the line of sight, as in the case of zero exposure (\ref{eq:is}), but the WFs will be different.

Assuming that the temporal evolution of the distorted wavefront is described by its translation (Taylor's \citeyearpar{Taylor} hypothesis of the frozen turbulence) by $w \tau$, we can describe the effect of temporal averaging by a convolution of the receiving aperture with a line of wind shift in the spatial domain \citep{FexpT2002,Martin1987}. In calculating the WF as integral of the filtered spatial power spectrum of scintillation, an additional multiplication by the temporal averaging filter appears.

\begin{figure}[h]
\centering
\psfig{figure=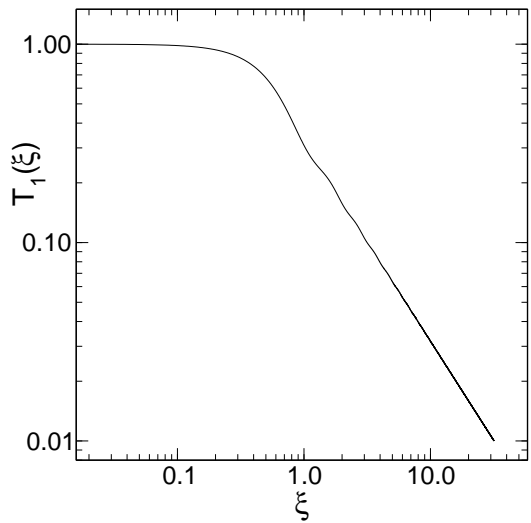,height=8cm}
\caption{ Wind shear spectral filter $A_s(w,\tau,f) = \mathcal T_1( f \tau w )$ \label{fig:tau1}}
\end{figure}

The new WFs $W'(w, \tau, h)$ will depend not only on the layer altitude $h$, as previously, but also on the wind speed $w = w(h)$ and averaging time $\tau$.
\begin{equation}
W'(w,\tau,h) = 9.62\,\lambda^{-2}\int_0^\infty {\rm d}f \, f^{-8/3} \sin^2(\pi\lambda h f^2) \,A(f)\, A_s(w,\tau,f),
\label{wdef}
\end{equation}
where $f$ is the modulus of spatial frequency, $A(f)$ is the aperture filter (axi-symmetric for the MASS apertures) and $A_s(w, \tau, f)$ is the additional spectral filter of the wind shear along the axis $x$, which is the square of the Fourier Transform of the rectangular window of the size $w \tau $. In polar coordinates it can be written as
\begin{equation}
A_s(w,\tau,f) = \frac{1}{2\pi}\int_0^{2\pi} {\rm d}\phi \, {\rm sinc}^2(f\tau w\cos\phi) = \mathcal T_1(f\tau w),
\end{equation}
where the function $\mathcal T_1(\xi)$ can be expressed through the Bessel functions $J_0$ and $J_1$ and the Struve functions $H_0$ and $H_1$ (the definition of Struve functions can be found in \citep{struve,abr}):
\begin{equation}
\mathcal T_1(\xi) = 2 J_0(2\pi\xi)-\frac{J_1(2\pi\xi)}{\pi\xi} - \pi J_0(2\pi\xi)H_1(2\pi\xi) + \pi J_1(2\pi\xi)H_0(2\pi\xi)
\end{equation}
The behavior of this function is shown in Fig.~\ref{fig:tau1}. Although the paper of \citet{FexpT2002} does not provide the analytical expression of this function, its correct asymptotic was given. We only refine the interval of the asymptotic. For small $\xi$, the function $\mathcal T_1(\xi)\approx 1 - \pi^2\xi^2/6$, as follows from its series expansion in the neighborhood of 0. The quadratic approximation provides accuracy better than 0.02 until the $\pi\xi<1$. When $\xi\to\infty$, the function $\mathcal T_1(\xi)\approx 1/\pi\xi$, and starting from $\xi\approx 1$ the relative difference is less than 0.04.

The asymptotic behavior of the function at $\xi\to 0$ and $\xi\to\infty$ can be used to describe the two limiting cases: the short exposure and long exposure regimes. The regime of short exposure $\tau w \ll D$ have been analyzed in detail by \citet{FexpT2002}.

\section{Long exposure regime}

Classical astronomical photometry almost always works in the approximation of the long exposure $\tau w \gg D$. In this situation, $f \tau w \gg 1 $ and one can use the asymptotics $\mathcal T_1(\xi) \approx 1/\pi\xi$. Then the WF looks like
\begin{equation}
W'(w,\tau,h) = 9.62\,\lambda^{-2}\frac{1}{\pi}\frac{1}{w(h)\tau}\int_0^\infty {\rm d}f \, f^{-11/3} \sin^2(\pi\lambda h f^2) \,A(f) = \frac{U'(h)}{w(h)\tau},
\label{winf}
\end{equation}
where $U'(h)$ is the value of the integral over spatial frequency. Note that the integrand has no singularities and is integrable without any problems, as well as in the zero-exposure case. If $f \to 0$, it tends to 0, justifying the use of this approximation. It is evident that the WF $W'(w, \tau, h)$ has a simple structure and the scintillation power averaged over exposure $\tau$ is defined by the following expression:
\begin{equation}
\sigma^2(\tau)=\int_0^\infty \frac{C_n^2(h)}{w(h)\tau}\,U'(h)\,{\rm d}h
\label{eq:long1}
\end{equation}
In terms of MASS the method, this expression has a very convenient form that allows, in principle, to determine the vertical distribution of $C_n^2(h)/w(h)$ from measurements made in a long-exposure regime. Note that the approximation of long exposure is valid at lower values of $\tau$ for small apertures, such as in the MASS device.

In the limiting case of infinitely small aperture $U'(h) = 13.52\,\lambda^{-2/3}\,h^{4/3}$ can be obtained using the fact that $A(f)\equiv 1$ and changing the variables \citep{Roddier81}. For the opposite case of large aperture $D \gg r_F$ (Fresnel scale $r_F=\sqrt{\lambda h}$), the function $U'(h)$ can be calculated by replacing the Fresnel filter $\sin^2(\pi\lambda h f^2)$ with $(\pi\lambda h f^2)^2$, as done by \citet{Roddier81}. Turning to the dimensionless frequency $q = fD$, we find that $U'(h) = 10.66\,D^{-4/3} h^2$ or
\begin{equation}
\sigma^2(\tau)=10.66\,D^{-4/3}\tau^{-1} \int_0^\infty \frac{C_n^2(h)\,h^2}{w(h)}\,{\rm d}h
\label{eq:long2}
\end{equation}
If $D=1$ and $\tau=1$, this expression represents the index $S_3^2$ introduced in paper \citet{ken2006}.

In the case of long exposures, the dependence of scintillation on airmass $M_z$ (or zenith angle) is obvious: $\sigma^2(\tau) \propto M_z^3$.  However, in the formulas (\ref{eq:long1}) and (\ref{eq:long2}) $w$ is the component of the wind velocity perpendicular to the line of sight. Naturally, the wind speed is normally directed horizontally. If the azimuth of the wind differs from the azimuth of the line of sight by $\pm 90^\circ$, then the perpendicular component is simply equal to the absolute speed. If the wind direction coincides with the plane of the line of sight, then perpendicular component is $ w_\perp = w\cos z $ and the dependence of $\sigma^2(\tau)$ on the airmass increases to $M_z^4$. The dependence of the scintillation power on azimuth was described by \citet{Young69}.

\section{Direct way to calculate the $S_3$ index}

The value $S_3^2$ is the variance of the stellar scintillation measured at 1~m telescope with 1~s exposure. The main point of the method for calculating the index $S_3$ (needed for estimating the photometric accuracy on large telescopes) directly from the MASS data is that $S_3^2$ is represented as a linear combination of long-exposure scintillation indices at different apertures of the MASS device. Unlike the instantaneous indices $s^2$, they will be denoted as $\sigma^2$.

\begin{figure}
\centering
\psfig{figure=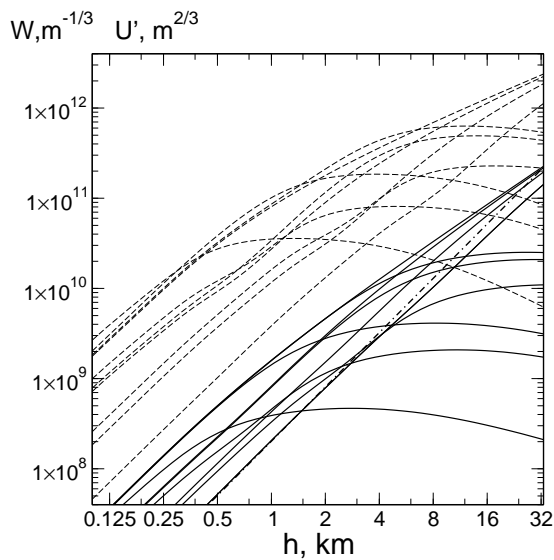,height=8cm}
\caption{ MASS WFs $W(h)$ (dashed) and $U'(h)$ (solid lines). Steady curves represent normal indices, saturated ones --- differential indices. Dot-dashed line shows a square-law altitude dependence \label{fig:up}}
\end{figure}

In the MASS data processing, the so-called atmospheric moments are calculated. In particular, the second moment is
\begin{equation}
M_2 = \int_0^\infty C_n^2(h)\,h^2 {\,\rm d}h,
\label{eq:m2}
\end{equation}
The problem of approximation of the $h^2$ dependence is solved in the MASS data processing by a linear combination of the WFs $W(h)$. The case under study here differs only in that the approximation of $h^2$ must be done with a set of $U'(h)$ functions, having the dimension of $m^{2/3}$. A set of $U'(h)$ is shown in Fig.~\ref{fig:up}. These functions are similar to the usual WFs, but are more smooth owing to the suppression of high spatial frequencies. It is clear that the time-averaged scintillation are much smaller. Therefore, the values of the functions $U'$ and $W$ differ by about 2 orders of magnitude. Both sets of WFs are computed for the polychromatic case with the spectral curve of the original MASS device and an A0\,V star.

The curve for the A aperture differs slightly from the WF of an infinitely small aperture at $\lambda = 0.5 \mu$m (effective wavelength). Strictly speaking, the limiting behavior for large apertures is also valid for the monochromatic case. For example, the real $U'(h)$ for the photometric band $V$ grows with $h$ more slowly. Moreover, telescope central obscuration also slows the growth of $U'(h)$. However, unlike the case of short exposure, these effects are small. Even for a 0.3-m aperture, the ratio of $U'(h)$ to $10.66\,D^{-4/3} h^2$ differs from 1 by no more than 5\% at 12~km altitude for the R band and with the central obscuration of 33\%. In the case of a 0.5-m telescope, all the differences do not exceed 2\% at all altitudes.

\section{Approximation of the quadratic dependence with the MASS WFs.}

\begin{figure}
\centering
\psfig{figure=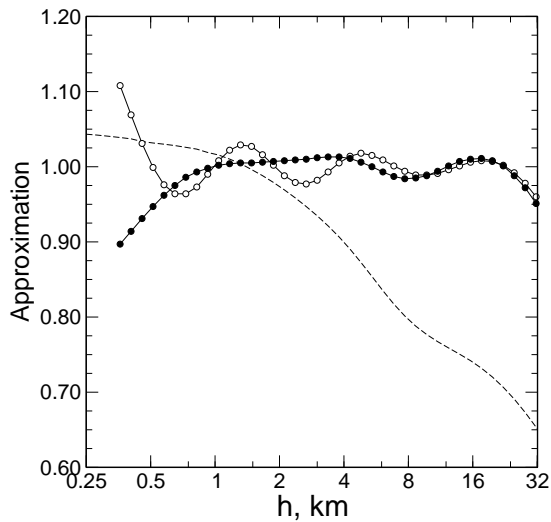,height=8cm}
\caption{ The ratio of the approximating linear combinations of the  WFs to $10.66\,h^2$. Black dots --- the approximation using only normal indices, empty circles --- using all 10 indices. The dashed  curve is the ratio of the WF for the D aperture of the device (11~cm) \label{fig:appr}}
\end{figure}

For a rough estimate of the $S_3$ index one can use an evaluation of $\sigma^2$ from measurements in the D aperture of the MASS device. Although its diameter is about 11~cm and the central obscuration is more than 50\%, the corresponding WF aproximates the quadratic dependence quite well. The maximum difference is achieved at high altitudes (see Fig.~\ref{fig:appr}) and is $\approx $25\% at the tropopause altitude. This means that the $S_3$ error will not exceed $\sim $15\%.

The accuracy of the quadratic approximation can be significantly improved by representing the WF of a 1~m telescope as a linear combination of functions $U'(h)$. In Table~\ref{tab:0} the coefficients of this expansion are listed. The coefficients $c_j$ are obtained by the decomposition on the whole set of $U'(h)$ functions, $d_j$ --- only on the 4 functions corresponding to normal indices in apertures A, B, C and D. Both approximations are shown in Fig.~\ref{fig:appr}, where the ratios of the approximated function to $10.66\,D^{-4/3} h^2$ are plotted. It is clear that both approximations provide a similar accuracy, better than 5\%, over the entire range of altitudes 0.5 -- 30~km, and better than 2\% in the
tropopause region. For the MASS/DIMM instrument the situation is slightly worse because the typical size of the aperture D is 8~cm and the central obscuration is larger.

In view of the linear relation between the indices and their corresponding WFs (\ref{eq:long2}) the following formula is correct:
\begin{equation}
S_3^2 = \sum_j c_j\sigma^2_j \quad\Bigl|\quad S_3^2 = \sum_j d_j\sigma^2_j\Bigr.,
\end{equation}
and the calculation the $S_3$ index reduces to the measurement of the variance of the flux in the MASS apertures averaged over 1~s.

\begin{table}
\caption{Approximation of $h^2$ by the WFs $U'(h)$ of the original
 MASS instrument. \label{tab:0}}
\centering
\bigskip
\begin{tabular}{lrrrrrrrrrr}
\hline\hline
  & A & B & C & D & AB & AC & AD & BC & BD & CD\ups \\[5pt]
\hline
$c_j$ \ups & -0.0200 & -0.0080 & 0.0160 & 0.0975 & 0.0057 & 0.0050 & -0.0060 & 0.0153 & 0.0106 & 0.0216 \\
$d_j$      & -0.0005 & -0.0090 & 0.0007 & 0.0907 & 0      & 0      & 0       & 0      & 0      & 0 \\[5pt]
\hline\hline
\end{tabular}
\end{table}

\section{Calculation of the $S_3$ index}

The required quantities are not calculated by the MASS software during measurements. However, all the necessary information is stored in the output files {\tt *.stm}, where the intensities in apertures A, B, C and D averaged over 1-s exposure are recorded. Note that these values in the file are given for the 1-ms exposure, in fact the counts are $10^4 \div 10^6$ and, consequently, the contribution of the photon noise is small.

Unfortunately, for a typical 1 minute acquisition time there are only 60 measurements. Therefore, the relative accuracy of the computed variance is approximately 25\%. For more accurate values, further averaging (over 4-minute interval) is needed.

Selection of valid data presents some problems because the fluctuations on 1-s temporal scale can include significant intensity changes of another origin --- due to changes in transparency, clouds, or large guiding errors. 

\begin{figure}
\begin{tabular}{cc}
\psfig{figure=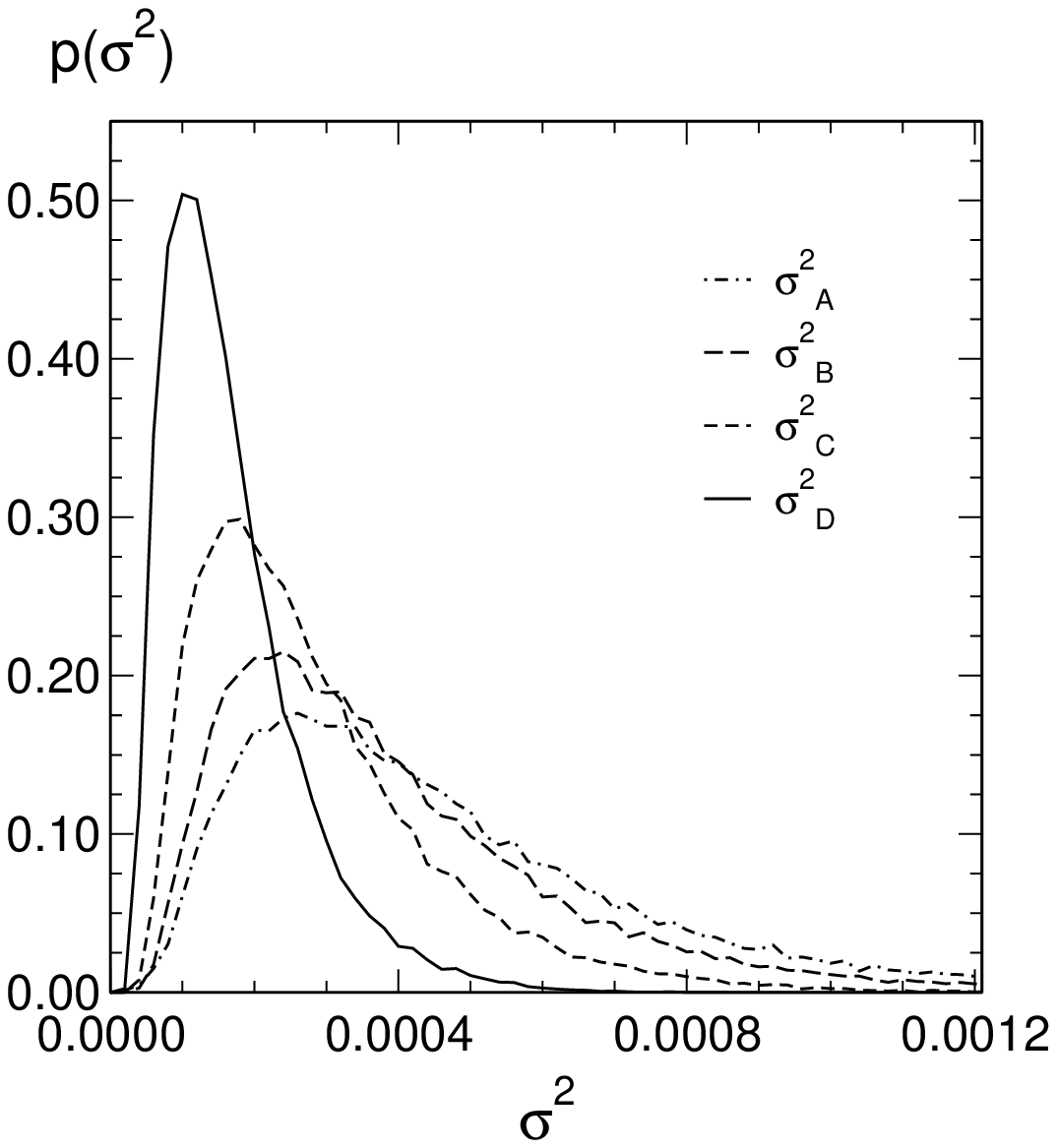,height=8cm}&
\psfig{figure=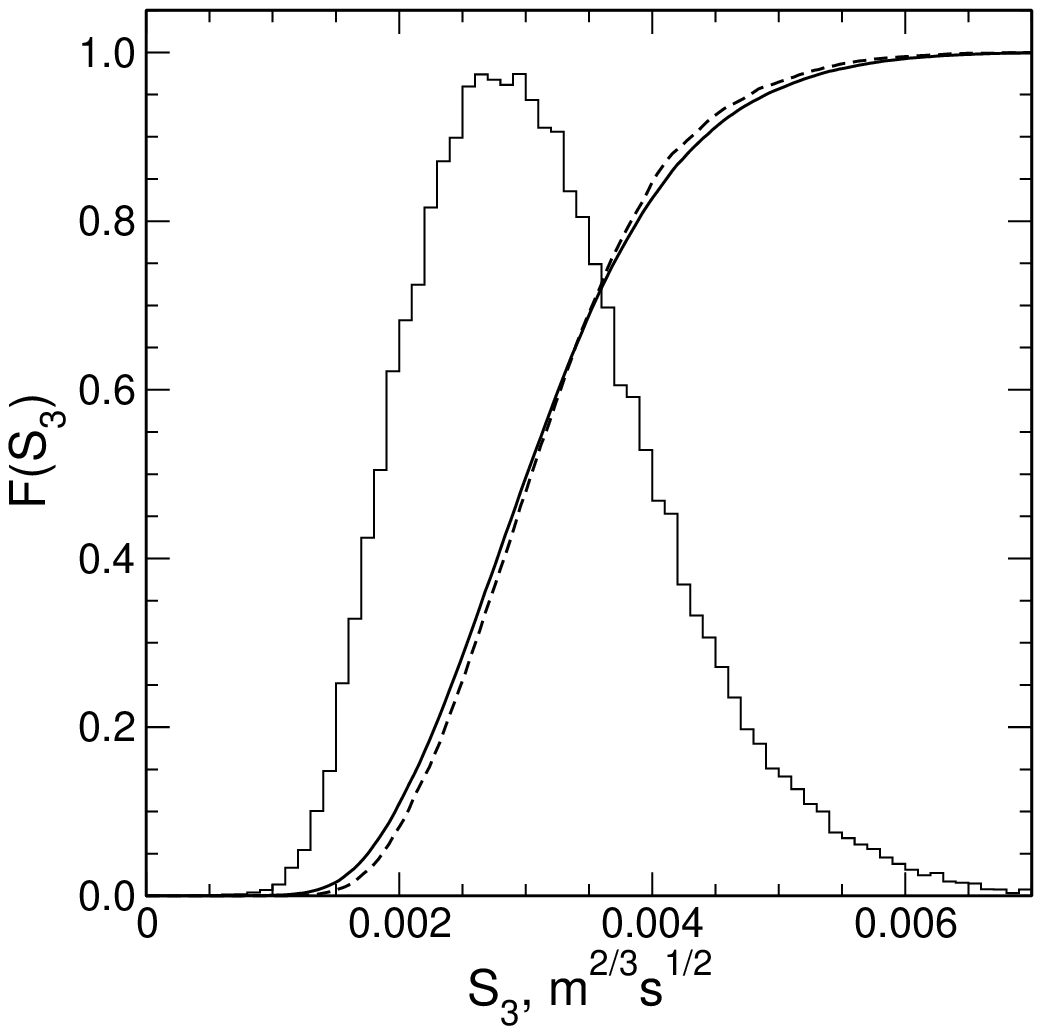,height=8cm}\\
\end{tabular}
\caption{Diffential distributions of the varinces of 1~s average
 fluxes in the A, B, C and D apertures of the MASS
 instrument \label{fig:distr-td}}
\caption{Distribution of the $S_3$ index. The solid line is obtained
 for variances estimated over 1 minute, dashed line --- over 4
 minutes. The stair-like line is the diffential distribution of
 $S_3$ \label{fig:distr-s3}}
\end{figure}

To investigate the atmospheric limit of photometric accuracy, the data of the measurements performed on the mountain Maidanak in 2005 -- 2007 with the MASS instrument were used \citep{maid2005}. The variances
$\sigma^2$ for all apertures were calculated with help of the mean square difference of adjacent 1-s means
\begin{equation}
\sigma^2 = \frac{1}{2(K-2)}\frac{1}{\bar F^2}\sum_{i=0}^{K-1}(F_i-F_{i+1})^2 - 0.001/\bar F,
\label{sdif}
\end{equation}
where $\bar F$ is the mean of the 1-s intensities $\{F_i\}$ in a given aperture. The last term is the contribution of the photon noise. 

Use of the differences effectively suppresses the contribution of intensity variations on time scales longer than 1 sec, of non-scintillation nature. At these scale, the scintillation itself is uncorrelated, which leads to the asymptotics $\sigma^2 \propto \tau^{-1}$. We used only data where the average count in the channel D was greater than 100 pulses/ms and the relative variance did not exceed 0.002.

Differential distribution of thus defined scintillation indices for the entire campaign period (45\,714 1-min points, or about 80\% of all measurements) are shown in Fig.~\ref{fig:distr-td}. Values are
reduced to the zenith. Medians and quartiles of the distributions of 1-s indices in the apertures A, B, C and D are given in Table.~\ref{tab:2}. The shape of the differential distributions is as expected and shows that data censoring does not distort them.

\begin{table}
\caption{Characteristics of the distributions of the variances 1~s  mean fluxes and the indices $S_3^2$ and $S_3$ \label{tab:2}}
\bigskip
\centering
\begin{tabular}{lrrrrrr}
\hline\hline
\ups               &      A   &      B   &     C    &     D    &      $S_3^2$         & $S_3$  \\[5pt]
\hline
quartile 25\% \ups & 0.000246 & 0.000208 & 0.000152 & 0.000088 & 5.83$\cdot 10^{-6}$  & 0.00242 \\
median             & 0.000386 & 0.000322 & 0.000234 & 0.000135 & 9.07$\cdot 10^{-6}$  & 0.00302 \\
quartile 75\%      & 0.000589 & 0.000488 & 0.000351 & 0.000201 & 13.65$\cdot 10^{-6}$ & 0.00370 \\[5pt]
\hline\hline
\end{tabular}
\end{table}

The $S_3^2$ index was calculated using the coefficients $d_j$ from Table.~\ref{tab:0}. The median value of $S_3^2$ is $0.91\cdot 10^{-5}\mbox{ m}^{4/3}\mbox{s}$. The index computed over 4- minute intervals has a slightly higher median of $0.93\cdot 10^{-5}\mbox{ m}^{4/3} \mbox{s}$. The difference is negligible, proving that the accuracy of the variance estimation over 1~minute accumulation time is sufficient. Clearly, this accuracy is mainly defined by the error of the variance in the D aperture. Cumulative distribution of the index $S_3$ is shown in Fig.~\ref{fig:distr-s3}. Naturally, this distribution is narrower than the distribution of $S_3^2$. The quartiles of this distribution are listed in Table.~\ref{tab:2}.

The median of the $S_3$ distribution equals $0.0030\mbox{ m}^{2/3}\mbox{s}^{1/2}$. This value is in good agreement with the estimation of $S_3$ for the Cerro Pachon and Cerro Tololo observatories from the paper \citet{ken2006}, where the medians of $0.0028$ and $0.0032\mbox{ m}^{2/3}\mbox{s}^{1/2}$, respectively, are given.

\begin{figure}[h]
\center
\psfig{figure=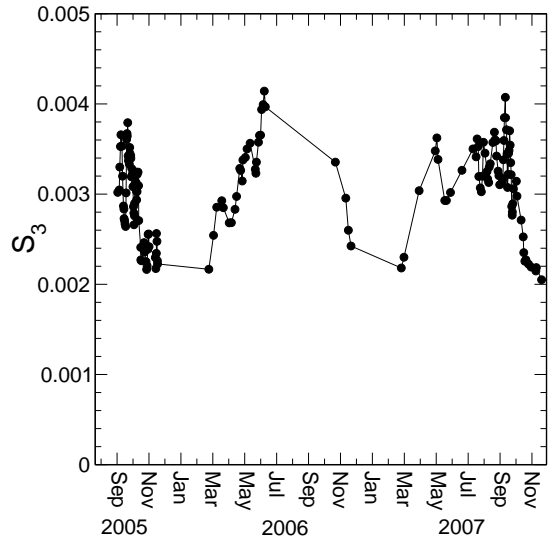,height=8cm}
\caption{Seasonal behavior of the median values of the index $S_3$\label{fig:s3-season}}
\end{figure}

The evolution of the quantity $S_3$ during the 2005 -- 2007 campaign is shown in Fig.~\ref{fig:s3-season}, where the medians over each $\approx $250~min data points are plotted. The annual cycle is
clearly visible. Seasonal variations are considerable, the index $S_3 $ varies by almost two times. The best photometric accuracy is expected in the period from October to April.

\section{Evaluation of the effective wind speed}

If we compare the formulas (\ref{eq:long2}) and (\ref{eq:m2}), we can see that their ratio gives some kind of estimation of the wind speed $\bar{\bar w}$. More precisely, the ratio of the second atmospheric
moment $M_2$ to the index $S_3^2$ gives the average value of $1/w(h)$ weighted by $C^2_n(h)\,h^2$:
\begin{equation}
\bar{\bar w} = \left\langle \frac{1}{w(h)}\right\rangle^{-1} = \frac{10.66\cdot M_2}{S_3^2}.
\label{eq:estim}
\end{equation}
The $M_2$ is calculated in the standard MASS data processing and, apart from a coefficient, equals the index $S_2^2$ introduced in \citep{ken2006}. The weight $C^2_n(h)\,h^2$ reaches maximum usually at
altitudes above 16~km and the expression (\ref{eq:estim}) evaluates thus high-altitude wind. Additional uncertainty is associated with averaging of reciprocals, when the resulting average may be underestimated. At Maidanak, the height of 16~km above the summit corresponds to the barometric altitude of 18 -- 19~km, or the pressure level of $\approx 70$~mbar.

\begin{figure}
\center
\begin{tabular}{cc}
\psfig{figure=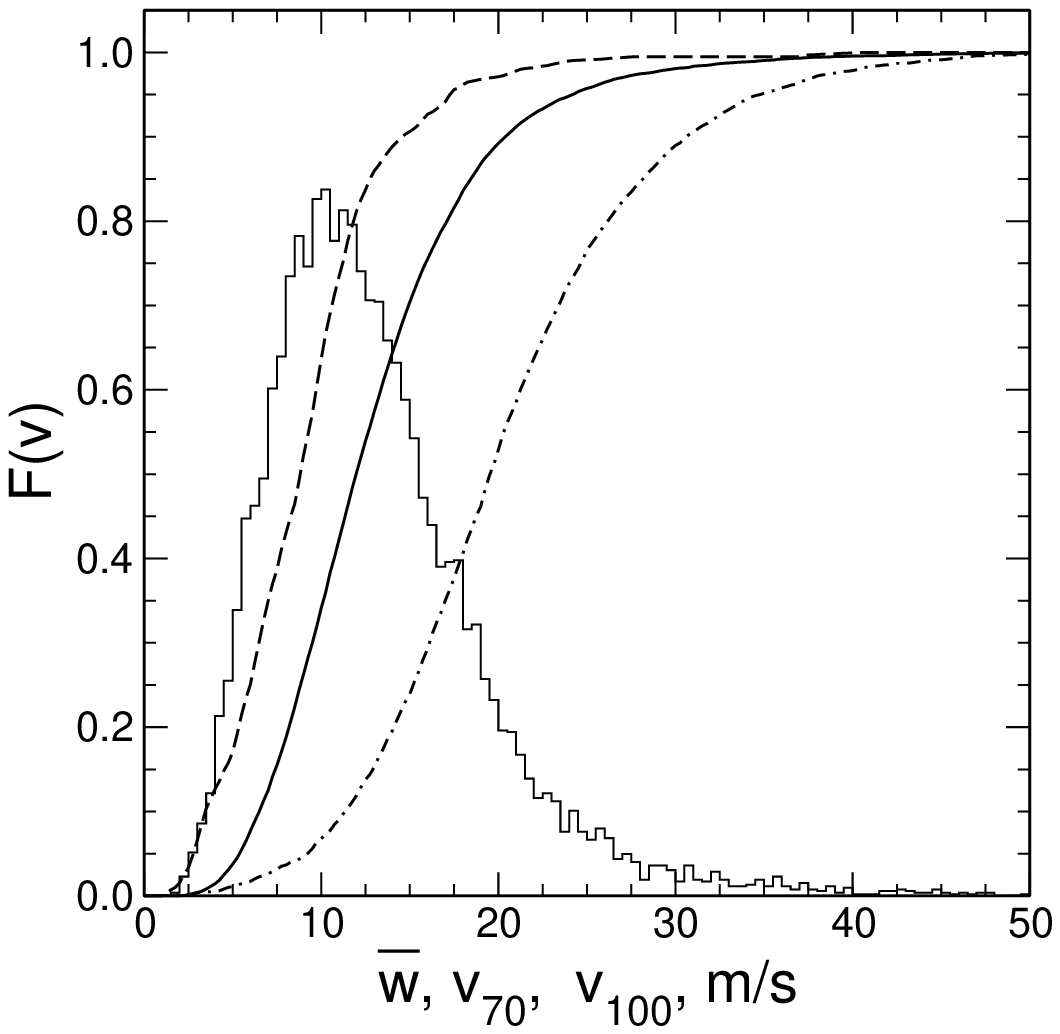,height=8cm} &
\psfig{figure=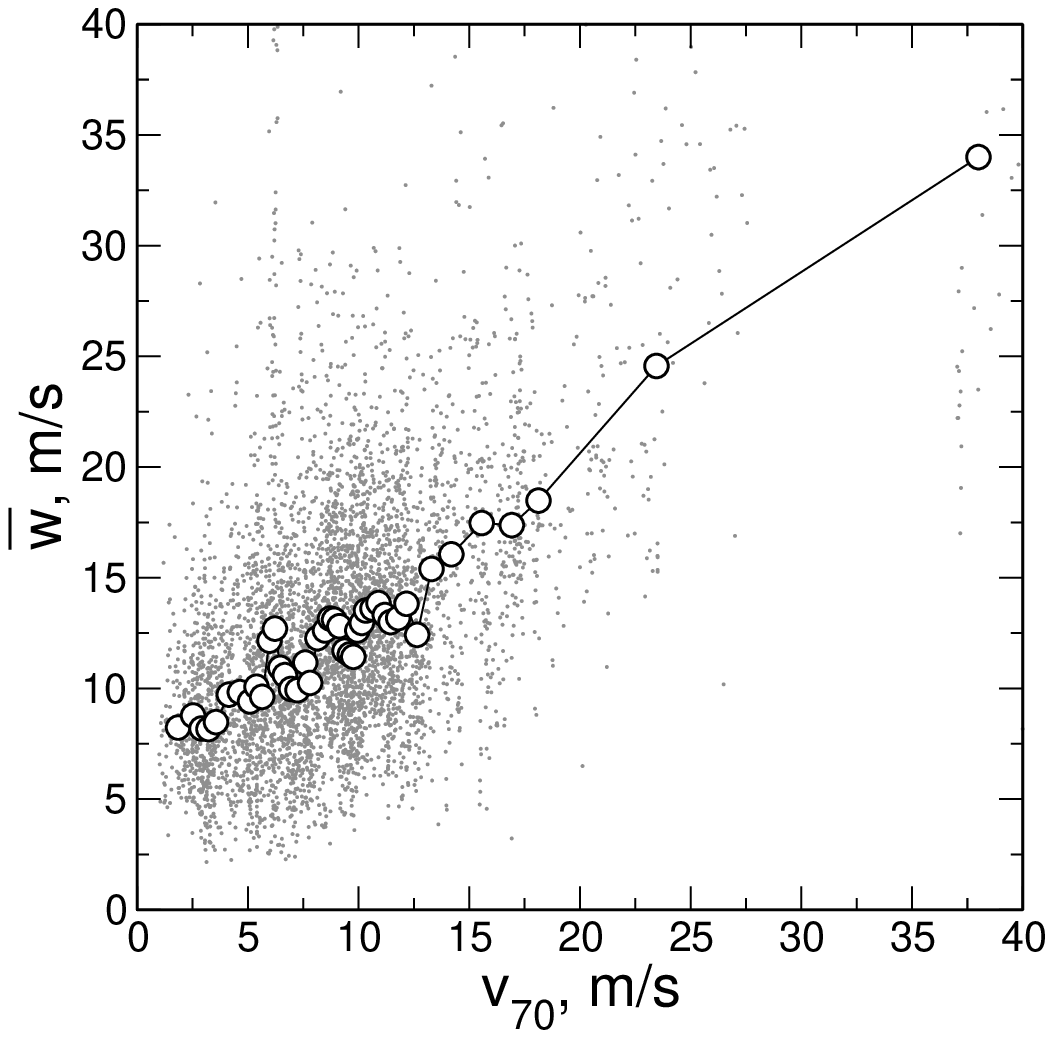,height=8cm}\\
\end{tabular}
\caption{Cumulative distributions of the winds. Solid line --- wind evaluated $\bar{\bar w}$ by formula (\ref{eq:estim}), dash-dotted --- wind at 100~mbar pressure level, dashed --- at 70~mbar level selected for moments of the measurements. Stair line --- differential distribution of the $\bar{\bar w}$ \label{fig:distr-v}}
\caption{Relation between the $\bar{\bar w}$ and wind at 70~mbar pressure level. Empty circles are meadian values in 250 points groups\label{fig:compar-v}}
\end{figure}

The result of high-altitude wind speed calculation are shown in Fig.~\ref{fig:distr-v}. The distributions of the wind speed $v_{70}$ and $v_{100}$ at pressure levels of 70 and 100~mbar as retrieved from the database NCEP/NCAR\footnote{http://www.esrl.noaa.gov/psd/data/reanalysis/reanalysis.shtml}
for moments of the measurements are also plotted. It is seen that the distribution of $\bar{\bar w}$ is bracketed by these two distributions, indicating the effective height determined by the $C^2_n(h)\,h^2$ weighting is indeed somwhere between these two levels. 

Comparison of the individual points of $\bar{\bar w}$ calculated for the values of $S_3^2$ and $M_2$ estimated over 4-min. with $v_{70}$  interpolated at these moments is shown in Fig.~\ref{fig:compar-v} in the form of a plot of the median values over groups of 250 points. A significant correlation and an almost linear relation between $\bar{\bar w}$ and $v_{70}$ are clearly visible. In the weak wind domain a systematic excess of $\bar{\bar w}$ over $v_{70}$ is observed. This may be due to the fact that in such situations the contribution of lower layers with faster wind and stronger turbulence becomes important.
 
\begin{figure}
\centering
\psfig{figure=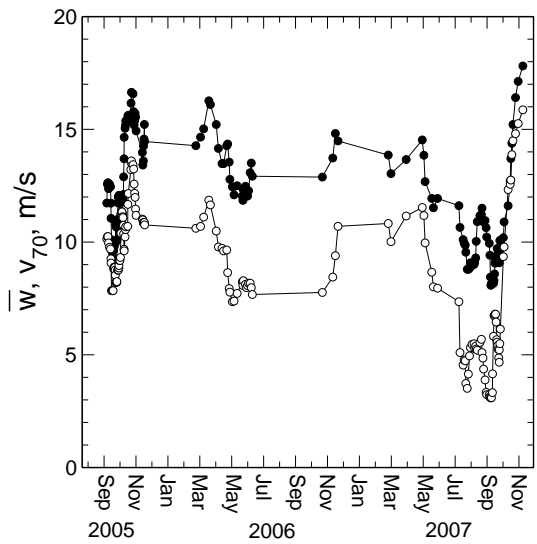,height=8cm}
\caption{Seasonal behavior of the wind $\bar{\bar w}$ (black dots) and  wind at 70~mbar $v_{70}$ (empty circles) for the entire period of  measurements  with  the  MASS  instrument  on  Maidanak \label{fig:season-wind}}
\end{figure}

In Fig.~\ref{fig:season-wind} the long-term behavior the of high-altitude wind speed above the mountain Maidanak is shown. It is seen that the minimum value of high-altitude wind $\bar{\bar w} \approx 10\mbox{ m/s}$ is observed from early July to mid-October. A similar behavior is characteristic of $v_{70}$, but not of wind at the pressure level of 200~mbar.

A combination of the WFs $U'(h)$ that grows with $h$ slower than $h^2$ (e.g. linearly) can be found. However, since the WFs of the normal scintillation indices increase with altitude faster than $h^{4/3}$, such approximation causes a noticeable increase in the noise. It is preferable to use the temporal index in the A aperture ($\propto h^{4/3}$) and the atmospheric moment $M_{4/3}$, computed from the indices $s^2$. In any case, the dominant layer will be located lower than in the case considered above and the ratio similar to the formula  (\ref{eq:estim}) is likely to reflect the wind speed in the tropopause.

\section{Conclusion}

A method to obtain statistical prediction of photometric accuracy based only on the data from the MASS instrument is proposed. Comparison of the results obtained by this method and by integration
of the OT and wind profiles \citep{ken2006} can be used to verify wind models above astronomical sites.

It is particularly interesting to perform such a comparison for the measurements with MASS instrument performed in Antarctica at Dome C, as described in \citep{ken2006}. There is some concern that their data
censoring procedure, leaving only 16\% of useful data, may bias the $S_3$ estimates. An ability to estimate wind speed at pressure level of 70 -- 100~mbar directly from the MASS data will help to check the applicability of the wind models and will avoid biases caused by replacing instantaneous values by their statistical estimates.

Additionally, we note here that 1-s averaging time in the case of MASS apertures is too long. A regime of the long exposure with the relevant asymptotics starts at shorter exposures on the order of 0.1 --- 0.2~s, and the use of such short time would markedly improve the accuracy and reliability of the temporal index. The behavior of WFs $U'(h)$ (Fig.~\ref{fig:up}) indicates that by using a set of precise measurements of temporal indices and the restored $C_n^2(h)$ profile, a reasonable solution of the system equations like (\ref{eq:long1}) for the unknowns $1/w(h)$ may be achieved.

Realization of this possibility will permit 1) to obtain additional information about the current state of the turbulent atmosphere, and 2) to correct more precisely the MASS/DIMM data for the effects of finite exposure time.

The author thanks his colleagues who participated in the campaign to measure the optical turbulence above the mountain Maidanak in 2005 -- 2007, producing the data used here. The work was partially supported by the RFBR grant 06-02-16902a.

\end{document}